\documentclass{aa}
\usepackage{epsfig}

\def\etal{{\rm et al.\thinspace}}
\def\eg{{\rm e.g.\ }}

\def\ie{{\rm i.e.\ }}
\def\cf{{\rm cf.\ }}

\def\spose#1{\hbox to 0pt{#1\hss}}
\def\ltsimm{\mathrel{\spose{\lower 3pt\hbox{$\sim$}}
	\raise 2.0pt\hbox{$<$}}}
\def\gtsimm{\mathrel{\spose{\lower 3pt\hbox{$\sim$}}
	\raise 2.0pt\hbox{$>$}}}

\def\km{{\rm\thinspace km}}
\def\cm{{\rm\thinspace cm}}
\def\s{{\rm\thinspace s}}

\def\massh{{\rm\thinspace m_{H}}}
\def\kmps{\hbox{${\rm\km\s^{-1}\,}$}}
\def\erg{{\rm\thinspace erg}}
\def\ergps{\hbox{${\rm\erg\s^{-1}\,}$}}
\def\Msol{\hbox{${\rm\thinspace M_{\odot}}$}}

\def\ergpcm2ps{\hbox{${\rm\erg\cm^{-2}\s^{-1}\,}$}}

\begin{document}
   
\title{The Formation of Broad Emission Line Regions in Supernova-QSO Wind
Interactions}

\author{J.M. Pittard \inst{1}, J.E. Dyson \inst{1},
S.A.E.G. Falle \inst{2} and T.W. Hartquist \inst{1}}

\institute{Department of Physics \& Astronomy, The University of Leeds, 
        Woodhouse Lane, Leeds, LS2 9JT, UK
\and
Department of Applied Mathematics, The University of Leeds, 
        Woodhouse Lane, Leeds, LS2 9JT, UK\\}

\offprints{J. M. Pittard, \email{jmp@ast.leeds.ac.uk}}

\date{Received date / Accepted date}

\abstract{
We show that a cooled region of shocked supernova ejecta forms in a 
type II supernova-QSO wind interaction, and has a density, an ionization
parameter, and a column density compatible with those inferred for the
high ionization component of the broad emission line regions in QSOs. 
The calculations are based on the 
assumption that the ejecta flow is described initially by a similarity
solution investigated by Chevalier (\cite{C1982}) and Nadyozhin 
(\cite{N1985}) and is spherically symmetric. Heating and cooling 
appropriate for gas irradiated by a nearby powerful continuum source is
included in our model, together with reasonable assumptions for the
properties of the QSO wind. The model results are also in agreement with
observational correlations and imply reasonable supernova rates.
}

\titlerunning{Formation of BELR in Supernova-QSO Wind Interactions}
\authorrunning{Pittard, Dyson, Falle \& Hartquist}

\maketitle

\label{firstpage}

\section{Introduction}
\label{sec:intro}
In active galactic nuclei (AGN), the primary physical mechanism for 
the excitation of the broad emission line regions (BELR) is 
photoionization by the underlying broad-band continuum
(\eg Osterbrock \& Matthews \cite{OM1986}; Clavel \etal \cite{C1991}). 
Evidence for at least a two-component 
structure of the BELR has been provided by \eg Collin-Souffrin \etal 
(\cite{CDT1982}, \cite{CDJP1986}) and Wills \etal (\cite{WNW1985}). 
One component can be identified as consisting of high ionization lines,
including Ly$\alpha$, C{\sc iii}], C{\sc iv}, He{\sc i}, He{\sc ii},
N{\sc v}, and other multiply ionized species, and is known as the HIL. 
The second component can be identified with the low
ionization lines which include the bulk of the Balmer lines, and lines of
singly ionized species (\eg Mg{\sc ii}, C{\sc ii} and Fe{\sc ii}), 
and is known as the LIL. The regions emitting
the LIL and HIL display different kinematics, as deduced from studies of the
profiles and line widths (\eg Gaskell \cite{G1988}, Sulentic \etal 
\cite{SMDCM1995}). In quasars, the HIL are also systematically 
blue-shifted with respect to the LIL (see various papers in Gaskell \etal
\cite{GBDDE1999}). Discussion of the BELR parameters
for a range of AGN properties can also be found in the articles in Gaskell
\etal \cite{GBDDE1999}.

Many theoretical explanations have been proposed for the origin of
the BELR. They include: i) magnetic acceleration of clouds off accretion 
discs (Emmering \etal \cite{EBS1992}); ii) cloud formation 
in shocks produced by the interaction of an accretion disc wind with a 
nuclear wind (Smith \& Raine \cite{SR1985}); iii) the interaction 
of an outflowing wind with the surface of an accretion 
disc (Cassidy \& Raine \cite{CR1996}); iv) interaction of stars with 
accretion discs (Zurek \etal \cite{ZSC1994}); v) 
tidal disruption of stars in the {\mbox gravitational} field of the BH 
(Roos \cite{R1992}); vi) interaction of an AGN wind with supernovae and 
star clusters (Perry \& Dyson \cite{PD1985}; Williams \& Perry 
\cite{WP1994}); and vii) emission from 
{\mbox accretion} shocks (Fromerth \& Melia \cite{FM2001}). Models
identified as containing serious difficulties include the
{\mbox formation} of BELRs by the thermal instability of a hot 
{\mbox optically} thin flow
(as \eg discussed by Beltrametti \cite{B1981} and Shlosman \etal 
\cite{SVS1985}). Perry \& Dyson (\cite{PD1985}) noted that this
{\mbox mechanism} will not occur when $L_{bol} > 10^{46} \ergps$ as Compton 
cooling dominates over bremsstrahlung, and therefore it cannot be
responsible for the observed BELR in high luminosity QSOs.
Two-phase equilibrium {\mbox models} (Krolik \etal \cite{KMT1981}) are 
also not applicable to BELRs in these sources because implausibly 
high values of the AGN mass-loss rate would be required 
(Perry \& Dyson \cite{PD1985}). The formation of the BELR in ionized red 
giant or {\mbox supergiant} winds (Scoville \& Norman \cite{SN1988}; 
Kazanas \cite{K1989}; Alexander \& Netzer \cite{AN1994}) has
difficulty in reproducing the observed broad line wings 
(Alexander \& Netzer \cite{AN1997}), and models involving the ballistic
deceleration of clouds have a number of problems, as summarized by
Osterbrock \& Matthews (\cite{OM1986}). There are also concerns about 
the {\mbox various} assumptions in the infalling and orbiting cloud models proposed 
by Kwan and Carroll (Kwan \& Carroll \cite{KC1982}; Carroll \cite{C1985}; 
Carroll \& Kwan \cite{CK1985}).

One model which accounts for the LIL emission 
was proposed by Collin-Souffrin \etal (\cite{CDMP1988}). In this model the
LIL emission arises from the surface of the accretion disc, which is 
illuminated by back-scattered X-rays from the central source. Typical electron
densities and absorption columns are inferred to be 
$n_{e} \gtsimm 10^{11}\;{\rm cm^{-3}}$ and 
$N_{H} \gtsimm 10^{24}\;{\rm cm^{-2}}$ respectively. Approximately
three quarters of the total luminosity of the broad-line emission is
estimated to arise in this fashion (Collin-Souffrin \etal \cite{CDMP1988}). 
The HIL, therefore, contribute about
one {\mbox quarter} of this emission. The characteristic mass of the
BELR in high luminosity QSOs is $\sim 100 \;\Msol$ 
(Osterbrock \cite{O1993}), and this is dominated by the HIL.

In this paper we look at one aspect of activity in AGN connected with stars,
and which is relevant to the work of Perry \& Dyson (\cite{PD1985}).
We investigate the interaction of supernova ejecta with 
the optically thin, low density, hot QSO wind, in the presence of 
intense continuum radiation. In particular we examine if 
shocked gas can radiate efficiently enough to 
cool to temperatures appropriate for the HIL. The evolution of SNRs in a 
high density static ambient medium has been previously studied by
Terlevich \etal (\cite{TTFM1992}), with particular 
application to the formation of BELRs in starburst models 
developed to obviate the existence of supermassive black holes in AGNs.
Although there are similarities between this work and ours, two
{\mbox differences} exist. First, the initial conditions 
for the supernova ejecta differ from those in our work. Second,
these authors did not include Compton cooling or any heating processes
in their calculations. These factors will influence the thermal evolution
of the shocked regions.

In Section~2 we discuss the use of a similarity solution to specify
initial conditions in the calculation and the cooling and heating rates
adopted. Section~3 contains {\mbox results} for the calculated evolution of a
remnant for each of several assumed sets of environmental conditions,
showing that the formation of a cool region of shocked ejecta having a
density, ionization parameter, and column density in harmony with those
inferred from observations occurs for reasonable assumptions. In 
Section~4 we summarize our conclusions and describe how the work can be
extended.

\section{Details of the calculations}
\label{sec:details}
Fits to the results of explosion models of type II supernovae indicate 
that power-law stratifications represent
adequate approximations to ejecta density and velocity profiles (\eg see
Woosley \etal \cite{WPE1988}, and earlier work by Chevalier \cite{C1976}
and Jones \etal \cite{JSS1981}), and have been {\mbox widely} used in 
analytical and numerical studies of remnant evolution. If
$\rho \propto r^{-n}$, these results indicate that $n \approx 12$ for the
high velocity ejecta.   
 
Self-similar solutions for the structure of the shocked ejecta 
and swept-up medium assuming power-law approximations for both of the
unshocked components are {\mbox derived} by Chevalier (\cite{C1982}) and 
Nadyozhin (\cite{N1985}), following earlier work by Parker (\cite{P1963}).
Their relevance to actual remnants was most recently highlighted by
SN~1993J in M81. High resolution, spatially resolved VLBI 
observations showed self-similar evolution of the azimuthally {\mbox averaged} 
radius (Marcaide \etal \cite{M1997}). From 
application of the Chevalier-Nadyozhin model,
the observations are best fitted with $n \approx 12$, in good agreement
with results from explosion models.

For $n \geq 3$, there must be an inner core of material with a shallower
density profile ($\rho \propto r^{-\delta}$) in order for the mass of 
the ejecta to be finite. Such a core 
can be seen in the results of explosion models (\eg Jones \etal 
\cite{JSS1981}, Suzuki \& Nomoto \cite{SN1995}). In the simplest case, 
which we adopt in this paper, a core with uniform density ($\delta=0$) is 
surrounded by a steep outer envelope ($n=12$). The speed of 
the core radius $v_{c}$, and the
density of the envelope $\rho_{e}(r)$ are then given by

\begin{eqnarray}
\label{eq:v_core}
v_{c} & = & \left[\frac{2(5-\delta)(n-5)E}{(3-\delta)(n-3)
M_{ej}}\right]^{1/2}, \\
\label{eq:rho_env}
\rho_{e} & = & g^{n} t^{-3} \left(\frac{r}{t}\right)^{-n},
\end{eqnarray}

\noindent where $E$ is the explosion energy, $M_{ej}$ is the ejecta mass,
and 

\begin{equation}
\label{eq:g_n}
g^{n} = \frac{1}{4 \pi (n-\delta)} \frac{[2(5-\delta)(n-5)E]^
{(n-3)/2}}{[(3-\delta)(n-3)M_{ej}]^{(n-5)/2}}
\end{equation}

\noindent (\cf Chevalier \& Fransson \cite{CF1994}). 
For $\delta=0$ and $n=12$ (which
we assume in this paper), the density of the core $\rho_{c}$ is given by

\begin{equation}
\label{eq:rho_core}
\rho_{c} = \frac{729}{1120 \pi} \left(\frac{3}{70} \frac{M_{ej}^{5}}{E^{3}}
\right)^{1/2} t^{-3}
\end{equation}

\noindent (\cf Band \& Laing \cite{BD1988}), whilst for $\delta=0$ and
variable $n$ the relative mass and energy of ejecta in the envelope 
compared to the core are (\cf Truelove \& McKee \cite{TM1999}):

\begin{eqnarray}
\label{eq:mcore_mej}
\frac{M_{c}}{M_{ej}} & = & \frac{n-3}{n}, \\
\label{eq:ecore_esn}
\frac{E_{c}}{E} & = & \frac{n-5}{n}.
\end{eqnarray}

\noindent Hence for $n=12$, 75 per cent of the mass and 58 per cent of the
explosion energy are in the core. 

If the explosion occurred in a pure vacuum, the structure of the ejecta
would evolve according to Eqs.~\ref{eq:v_core}-\ref{eq:rho_env}.
However, when the ejecta interact with a surrounding medium,
the steep envelope acts as a compressible piston,
and the radius of the core relative to the reverse shock increases with
time. As long as ejecta in the steep envelope continue to pass through 
the reverse shock the solution is self-similar, but once the core radius 
reaches this point the solution abruptly ceases this behaviour. 

\begin{figure}
\begin{center}
\psfig{figure=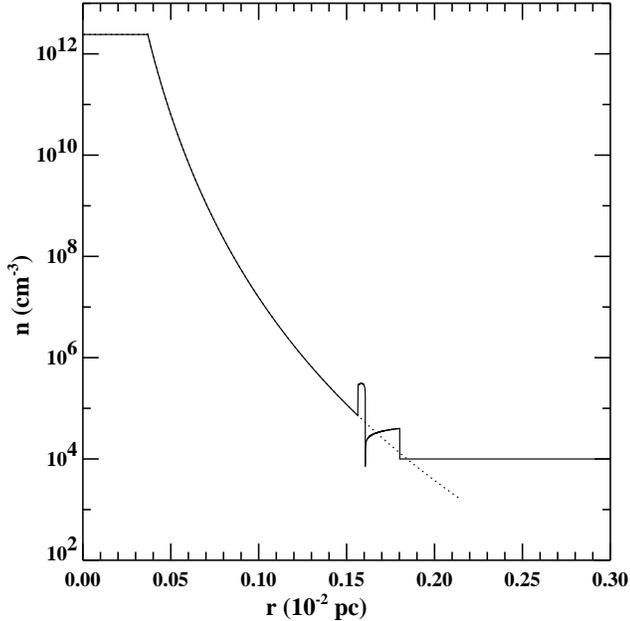,width=8.5cm}
\end{center}
\caption[]{The density profile for ejecta from a type II SN explosion, 
modelled as an $n=12$ power-law for
the envelope and a $\delta=0$ power-law in the core, an
explosion energy of $10^{51}$~erg and an ejecta mass of $10 \Msol$. The dotted 
line shows the solution given by Eqs.~\ref{eq:v_core}-\ref{eq:rho_env}
which is valid for ejecta expanding into a total vacuum. The solid line
shows the case when the progenitor is surrounded by a constant density 
medium with $n = 10^{4} \;{\rm cm^{-3}}$: the outer part of the ejecta 
and the swept-up ambient medium are compressed into the Chevalier-Nadyozhin 
similarity form.}
\label{fig:compress_piston}
\end{figure}

In this work we adopt the Chevalier-Nadyozhin {\mbox similarity} solutions
to specify our initial conditions, with the ejecta distribution being 
specified by an inner core with $\delta=0$ and an outer envelope with $n=12$. 
Whilst this is a simplification to actual distributions obtained from
explosion models, at this stage we aim to keep our results as general
as possible. It also has the benefit of being easily reproducable and 
scale-free. In all our calculations we assume a canonical explosion energy 
of $10^{51}$~ergs and ejecta 
mass of $10 \Msol$, which is typical of a SN of type~II. 
Fig.~\ref{fig:compress_piston} shows a typical density profile 
at $t=0.1$~yr.

Heating and cooling rates for a canonical AGN spectrum were kindly supplied 
by Tod Woods (\cf Woods \etal \cite{WKCMB1996}) and are included in our
calculation. We made use of an adaptive grid hydrodynamic code 
(see \eg Falle \& Komissarov \cite{FK1996}, \cite{FK1998}), which
is ideally suited to this problem where regions which contain
small-scale structure are located within much smoother regions.

\begin{figure*}
\begin{center}
\psfig{figure=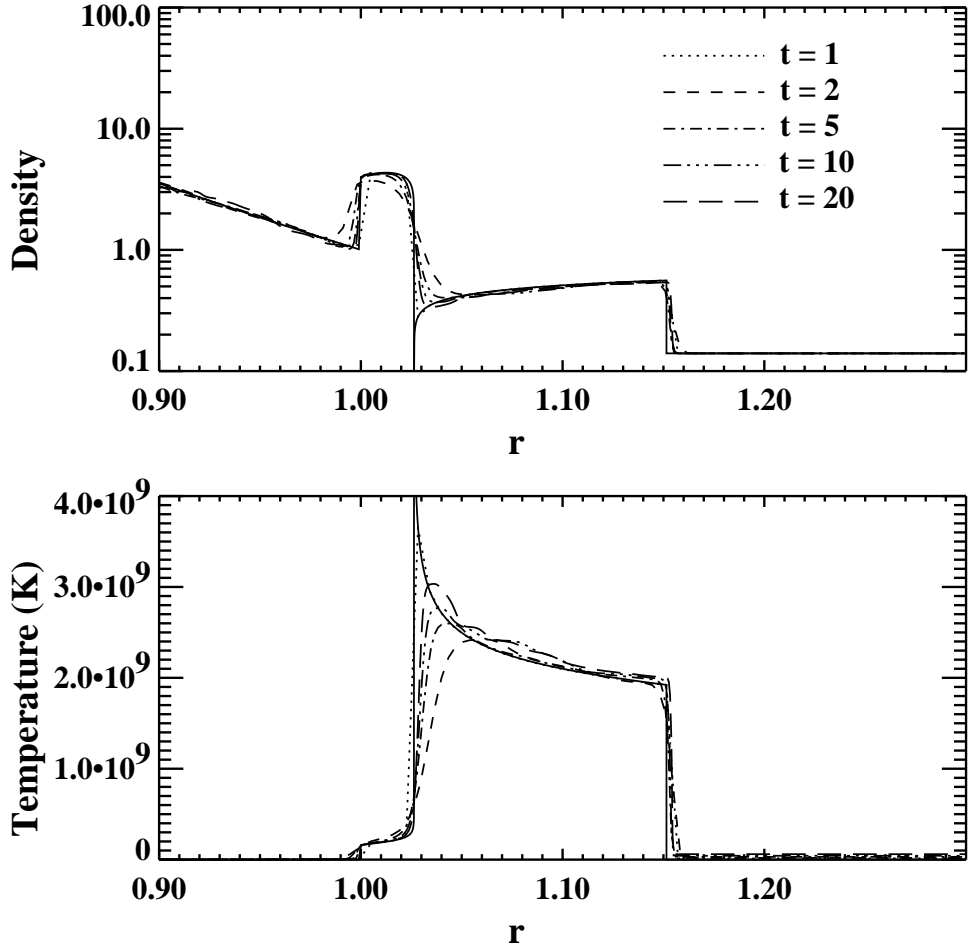,width=13.0cm}
\end{center}
\caption[]{Comparison of evolved radial profiles of density
and temperature rescaled against the similarity
solution (solid). The times are measured by the homologous expansion, with
$t=1$ corresponding to the initial mapping. The initial number of grid 
points in the region of
shocked ejecta was four. The curves are not coincident because of start-up
errors, finite numerical resolution, and unavoidable diffusion and 
dissipation in the numerical scheme. As the remnant expands the effective 
resolution of the 
calculation increases, and the profiles converge towards the original 
similarity solution.}
\label{fig:sim_advect}
\end{figure*}

To test the accuracy of the code we first imposed the similarity structure
on the flow and saw whether it sustained itself. The ambient medium had 
constant density, zero velocity, and negligible pressure. 
In Fig.~\ref{fig:sim_advect} we show the results of this 
test, which compare favourably with the results from other codes in the 
recent literature (\eg Blondin \etal \cite{BBR2000}).

The heating and cooling rates are tabulated as functions of 
temperature and ionization parameter, $\Xi$ ($= F/cp$ where $F$ is the 
local ionizing flux, $c$ is the speed of light, and $p$ is the gas 
pressure), and are valid in the {\mbox optically} thin, low-density regime. 
$\Xi$ is effectively a measure of the ratio of radiation pressure to the 
gas pressure, and is $2.3 P_{rad}/P_{gas}$ for fully ionized gas of cosmic
abundance. The heating rates include Compton and photoionization heating of 
all the ionization stages of hydrogen, helium, and some trace metals 
(particularly C, N, O, Si, S, and Fe). The cooling includes
collisional excitation, recombination cooling, 
Compton cooling, collisional ionization, and free-free losses. 
The gas is assumed to be free of dust (Laor \& Draine \cite {LD1993}; but
see review by Osterbrock \cite{O1993} for an excellent summary of our current
understanding of dust in AGN).
The rates were calculated using the {\sc CLOUDY} photoionization 
code, its standard AGN spectrum, and solar abundances. 
Whilst abundances in AGN remain a very contentious issue, there is
considerable evidence that the nuclear regions are not metal poor, even
at high redshift, and also little evidence that the metallicity of the
BELR changes with redshift (see Artymowicz \etal \cite{ALW1993}).
Though it seems certain that nitrogen is overabundant by factors of 
$\sim 2-9$, particularly for high-redshift sources (\eg Hamann \& 
Ferland \cite{HF1992}), most theoretical work has been based on the assumption
that the gas is of solar {\mbox composition}: we also apply this assumption.
More details of the heating and cooling rates can be found in 
Woods \etal (\cite{WKCMB1996}).

\begin{figure}
\begin{center}
\psfig{figure=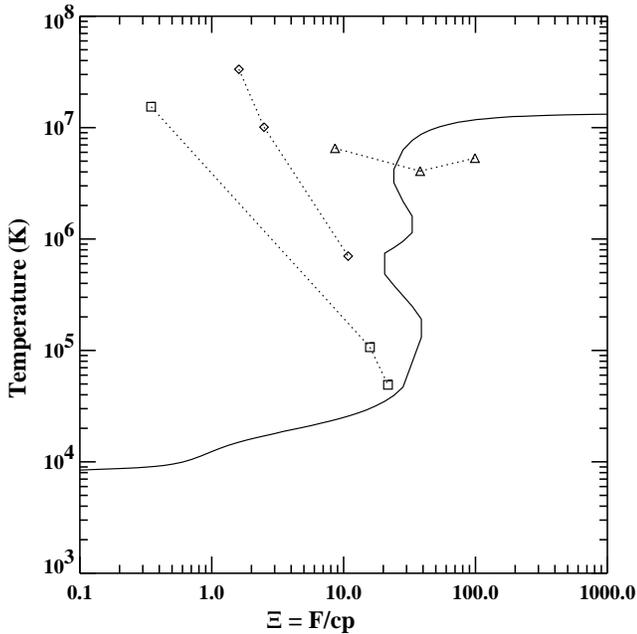,width=8.5cm}
\end{center}
\caption[]{Thermal equilibrium curve for the standard AGN spectrum in 
CLOUDY (see Woods \etal \cite{WKCMB1996}). The symbols refer to the
thermal evolution of the cool region in Models~A (diamonds), B (triangles) 
and C (squares) (Fig.~\ref{fig:clump_form}, Fig.~\ref{fig:clump_noform} and
Fig.~\ref{fig:clump_form_agn46_u3e8} respectively).}
\label{fig:teq}
\end{figure}

In Fig.~\ref{fig:teq} we show the thermal equilibrium curve for the 
assumed AGN spectrum. At low temperatures {\mbox photoionization} heating and 
cooling due to line excitation and recombination are in near balance.
At large {\mbox temperatures}, the equilibrium arises from a balance of Compton
heating and cooling. The solutions with {\mbox positive} slope, 
\ie $dT/d\Xi > 0$, are thermally stable (Field \cite{F1965}). If the slope 
is negative, solutions can still be {\mbox thermally} stable 
if the gas cools isochorically, 
whilst they are thermally unstable if the gas cools isobarically. Note that
there is a small range of ionization parameters for which
there are stable solutions at intermediate temperatures ($T \sim 10^{6}$~K).
The exact shape of this part of the thermal equilibrium curve
is poorly known and most likely varies substantially from source to source,
since it is a complicated function of the irradiating spectrum, the
assumed abundances and thermal processes (\cf Krolik \etal \cite{KMT1981}).
For a given object it is entirely possible that there are no multi-valued 
equilibrium temperatures for any $\Xi$. Therefore, in our discussion of 
the following {\mbox results} we do not assign too much importance to the 
exact behaviour of the simulations in this part of parameter space since we
are not modelling a specific object. In Sec.~\ref{sec:results} 
we further show that the cooling gas often does not
pass through this region of parameter space.

To obtain cool gas in thermal equilibrium we require ionization
parameters $\Xi \ltsimm 10$. 
As noted by Perry \& Dyson (\cite{PD1985}), shocked gas cooled back to
equilibrium can have a value of $\Xi$ much lower than its
pre-shock value. This is because $\Xi$ does not change if the gas cools
isobarically and the post-shock pressure is much greater than the
pre-shock value. Therefore strong shocks can create conditions
for the gas to cool to temperatures much lower than the surrounding
ambient temperature. 

The ionization parameter of the QSO ISM can be {\mbox written} as

\begin{equation}
\label{eq:ionp_wind}
\Xi_{w} = 2 \times 10^{6} \frac{L_{ion}}{L_{bol}} \frac{L_{47}}{n_{w} 
r_{pc}^{2} T_{8}},
\end{equation}

\noindent where $L_{ion}$ and $L_{bol}$ are respectively the 
ionizing and bolometric luminosity of the central source, 
$L_{47}$ is the {\mbox bolometric} luminosity of 
the central source in units of $10^{47} \ergps$, $n_{w}$ is the 
number density of the wind, $r_{pc}$ is the distance of the shock from
the central source in {\mbox units} of parsecs, and $T_{8}$ is the temperature
of the wind in units of $10^{8}$~K.

The ionization parameter of shocked gas cooled back to equilibrium 
with the radiation field, $\Xi_{shk}$, is given by

\begin{equation}
\label{eq:ionp_shk}
\Xi_{shk} = \frac{T_{8}}{4 \omega^{2}} \Xi_{w},
\end{equation}

\noindent where $\omega$ is the velocity of the pre-shock gas relative to the
shock in units of $0.01c$ (\cf Perry \& Dyson \cite{PD1985}). For 
supernovae in the central regions of QSO, this {\mbox velocity} is a 
combination
of the velocity of the blast wave, the {\mbox velocity} of the ambient medium,
and the Keplerian {\mbox velocity} of the progenitor star. The QSO medium 
has {\mbox characteristic} {\mbox velocities} of 
$v_{wind} \sim 1,000-3,000 \kmps$ 
($\omega = 0.3-1.0$, Williams \etal \cite{WBP1999}), whilst the expansion
velocities of young ejecta-dominated SNRs satisfying the 
Chevalier-Nadyozhin similarity solution can be greater than 
$10,000 \kmps$. Therefore, it is not difficult to obtain
$\Xi_{shk} \ltsimm 1$ for which $T_{eq} \sim 10^{4}$~K
(see Fig.~\ref{fig:teq}). 
The crucial question is whether the shocked gas remains at high
pressures long enough to cool from its post-shock
temperature to $T \sim 10^{4}$~K. This requires that the dynamical
timescale for the SNR is longer than the cooling timescale of the 
post-shock gas.

\section{Results}
\label{sec:results}

\subsection{Adiabatic evolution}
\label{sec:adiabatic_evol}
We initially investigate the adiabatic evolution of a 
SNR surrounded by a static, constant density medium with $n_{w} = 10^{6} 
\;{\rm cm^{-3}}$ and $T_{w} = 1.33 \times 10^{7}$~K. Whilst this value 
of $n_{w}$ is marginally optically thick to X-rays (at the Fe 
K-shell edge) and to Compton scattering for distances greater than 
$\approx 0.5$~pc, it is of no concern provided that the density
falls with radius on similar scales. However, as our results show,
the radius of the supernova {\mbox remnant} remains significantly smaller than this 
scale. We note also that our chosen ambient density is similar to that
found by Williams \etal (\cite{WBP1999}). In the central regions of 
AGN, the ISM density is primarily determined by mass-loss from the central
stellar cluster through winds and supernova {\mbox explosions}. 
Collisions and tidal disruptions are {\mbox negligible} except 
in the central regions of the densest
{\mbox clusters}. Compared to the central cluster, the mass flux from the
external ISM (\eg flow into the nucleus from a galactic bar) is unlikely to
provide a significant fraction of the fuelling requirements of the AGN
as a {\mbox continuous} source (Shlosman \etal \cite{SBF1990}).
Our value of $T_{w}$ is also {\mbox characteristic} of the Compton 
temperature in a hard QSO radiation field.

The assumption of a static ambient medium
keeps the problem in its simplest form. For the unshocked ejecta we set
$T = 10^{4}$~K. The Chevalier-Nadyozhin similarity solution is derived
on the assumption of negligible thermal energy in the unshocked ejecta 
and ambient medium, and our specified temperature implies thermal 
pressures orders of magnitude below the relevant ram pressure.
Whilst this also implies exceedingly high absorption columns through
the ejecta core, the covering fraction is very small.
 
\begin{figure}
\begin{center}
\psfig{figure=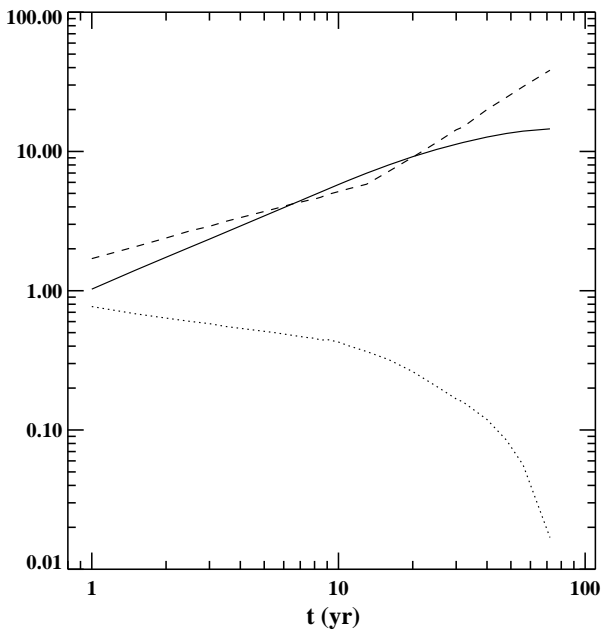,width=8.5cm}
\end{center}
\caption[]{Evolution of the radius (solid) and speed (dots) of the contact 
discontinuity of an adiabatic SNR as a function of time (arbitrary units). 
At $t=1$~yr, their respective values
are 1.0264 and 0.7698. The SNR evolves in a self-similar fashion 
with $r \propto t^{3/4}$ and $v \propto t^{-1/4}$ until
the reverse shock reaches the radius of the ejecta core. This occurs at 
$t \approx 9$~yr. Also shown is the time evolution of $\Xi$ 
(dashes), which at $t=1$~yr has 
a value of 1.7. Prior to $t \approx 9$~yr it evolves as $t^{1/2}$,
but increases in value much more rapidly for $t \gtsimm 13$~yr.}
\label{fig:snr_evol_nocool}
\end{figure}

In Fig.~\ref{fig:snr_evol_nocool} the radius and speed of the contact 
{\mbox discontinuity} are shown as functions of time. As discussed earlier, 
the solution remains self-similar until the reverse shock propagates to 
the edge of the core: this occurs at $t \approx 9$~yr, and is reflected in the
structure of the curves in Fig.~\ref{fig:snr_evol_nocool}.

Although we have not included heating and cooling terms in this 
calculation, it is instructive to assume a {\mbox radiation} field and to plot 
the evolution of the ionization parameter of the equilibrium temperature
of the shocked gas. This gives
an impression of whether the shocked gas is likely to heat or cool if
these terms are included. Assuming that the remnant is 0.33~pc distant
from a central ionizing source of $L_{ion} = 10^{47} \ergps$, we show
in Fig.~\ref{fig:snr_evol_nocool} the time evolution of this parameter 
in the shocked ambient medium. This position was chosen because, with the 
assumption 
of constant ionizing flux, $\Xi$ tracks the inverse of $p$ and the profile 
of $p$ is flatter at this point. Fig.~\ref{fig:snr_evol_nocool} shows that
$\Xi$ initially evolves as $t^{1/2}$, but increases more rapidly
once the flow diverges from self-similarity. This is due to a large
increase in the volume of the shocked region (as the reverse shock 
begins to move back towards the centre of the remnant in the Lagrangian 
frame) at roughly {\mbox constant}
total energy, which leads to a reduction in the internal energy per unit
volume and hence pressure, and the corresponding increase in $\Xi$. 
It is therefore apparent that the formation of cool clouds in a 
SNR shock is favoured at {\mbox early} times, when the pressure of the shocked
region is high and $\Xi$ is low. In particular, since $\Xi$ rapidly
increases once the ejecta core reaches the reverse shock, a cool
region has the best opportunity to form before this point.
Therefore we equate the dynamical timescale of the remnant, $t_{dyn}$, 
with the time at which the ejecta core reaches the reverse shock, and 
require that the shocked gas has a cooling time, $t_{cool}$, shorter than 
this. We derive expressions for $t_{dyn}$ and $t_{cool}$ in Appendix~A.

We also require that the ionization parameter $\Xi$ of the post-shock gas at
$t = t_{dyn}$ is low enough to allow the gas to cool to 
$T \approx 2 \times 10^{4}$~K. Cold gas will not form if 
$\Xi \gtsimm \Xi_{crit}$ (where $\Xi_{crit} \ltsimm 30$ for the
AGN spectrum used to generate the thermal equilibrium curve in 
Fig.~\ref{fig:teq}) whether or not $t_{cool} \ltsimm t_{dyn}$.
We derive an expression for this condition also in Appendix~A. Hence for cold
gas to form we need to satisfy $t_{cool} \ltsimm t_{dyn}$ and 
$\Xi)_{t_{dyn}} \ltsimm \Xi_{crit}$. Both of these conditions are evaluated
for each of the following models.

\subsection{Radiative evolution}
\label{sec:rad_evol}
We now include the full heating and cooling rates in
all of our remaining simulations. We start the evolution on the hydrodynamic 
grid at a time early enough for the dominant cooling of the shocked gas 
to be adiabatic expansion. As the remnant expands, 
radiative cooling gradually increases in importance, and a smooth
transition from the adiabatic self-similar solution into the radiative 
regime occurs. Because the post-shock temperatures can be very high early on 
($T \gtsimm 10^{9}$~K), we extrapolate the heating and cooling functions
used by Woods \etal (\cite{WKCMB1996}) with a second order polynomial.
This should be a reasonable approximation to the true rates because
over the range $10^{8} \ltsimm T \ltsimm 10^{9}$~K, cooling
and heating are dominated by Compton or bremmstrahlung processes which are 
{\mbox smoothly} varying functions of $T$. Furthermore, since 
the cooling of the 
remnant is initially dominated by adiabatic expansion, the inaccuracies
in the correct rates from this extrapolation are small. Finally, as
the remnant expands, the pre-shock ejecta are slowly heated from 
$T = 10^{4}$~K to $\sim 10^{6}$~K. We assume again that the remnant
is 0.33~pc distant from a central source with $L_{ion} = 10^{47} \ergps$.
The parameters for this model (Model~A) are listed in Table~\ref{tab:param}.

\begin{table}
\begin{center}
\caption{Parameters for the models considered in this paper. All models
have the same ionization parameter and temperature for the ambient medium
($\Xi \approx 150$, $T_{w} = 1.33 \times 10^{7}$~K).}
\label{tab:param}
\begin{tabular}{lllll}
\hline
Model & $n_{w}$ & $v_{w}$ & $F_{ion}$ & Cool \\
 & (${\rm cm^{-3}}$) & $(\kmps)$ & ($\ergpcm2ps$) & Regions \\
\hline
A & $10^{6}$ & 0 & $7.67 \times 10^{9}$ & Y \\
B & $10^{4}$ & 0 & $7.67 \times 10^{7}$ & N \\
C & $10^{4}$ & 3000 & $7.67 \times 10^{7}$ & Y \\
\hline
\end{tabular}
\end{center}
\end{table}

\begin{figure*}
\begin{center}
\psfig{figure=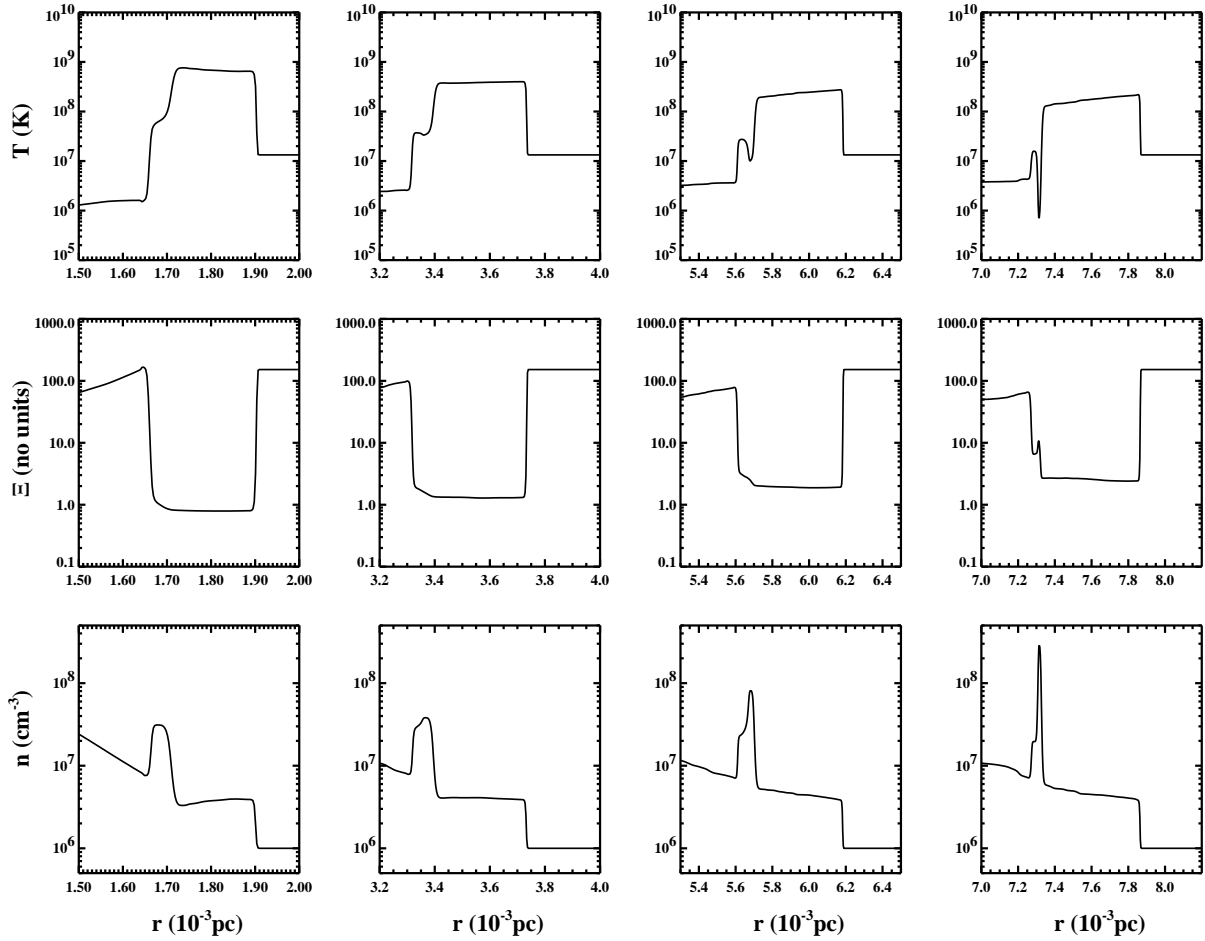,width=16.0cm}
\end{center}
\caption[]{The evolution of the remnant of a type-II supernova for
a constant density medium with $n_{w} = 10^{6} \;{\rm cm^{-3}}$ and 
$\Xi_{w} \approx 150$ (Model~A). The medium is stationary with 
respect to the remnant, and the entire area is immersed in a radiation field 
appropriate to an AGN of ionizing luminosity $10^{47} \ergps$ with the central
engine at a distance of 0.33~pc. The panels show from top to bottom the
evolution of the temperature, ionization parameter, and density of the 
region of hot shocked gas. This is bounded on its right edge by the forward
shock propagating into the ambient medium and on its left edge by the
reverse shock. From left to right the corresponding times are
0.2~yr, 0.5~yr, 1.0~yr and 1.4~yr. The formation of a cooled region of gas
can clearly be seen.}
\label{fig:clump_form}
\end{figure*}

In Fig.~\ref{fig:clump_form} the evolution of the shocked region is shown.
We find that the shocked {\em ejecta} substantially cool with heating and 
cooling rates {\em included} in the calculation. 
By $t=1.0$~yr a cool region has formed with a temperature 
less than that of the undisturbed ambient medium. At $t=1.4$~yr
it has a temperature of $7.0 \times 10^{5}$~K. Divergence from
the adiabatic self-similar solution can be seen in the time evolution
of the $T, \Xi$, and $\rho$ profiles. In the {\mbox temperature} profiles, 
this is first manifested as a change in the slope of
the region of shocked ambient material from $dT/dr > 0$ to
$dT/dr < 0$. The formation of the cool region at later times is of course 
a much larger divergence from the self-similar solution. 
In the density profiles, the formation of the cool region is
apparent as a sharp coincident growth in density. The profiles of 
ionization {\mbox parameter} show the gradual temporal increase 
in $\Xi$ expected 
from the evolution of the self-similar solution. However, 
at $t=1.4$~yr, the value of $\Xi$ associated with the cool material is 
substantially larger than the value of the immediate surroundings. 
This is due to the radiative losses becoming so high that cooling 
no longer takes place isobarically. In the limit that the cooling 
timescale, $t_{c}$, is much less than the appropriate dynamical 
timescale, $t_{d}$, the cold gas would cool isochorically. Here
$t_{d}$ is the timescale for the hot post-shock gas to respond to the
rapid depressurization of this material, and is of the order of the 
timescale for
collapse of the reverse shock, $\approx l/(v_{cd}-v_{rs})$, where
$l$ is the length scale of the shocked ejecta, and $v_{cd}$ ($v_{rs}$) is the
velocity of the contact discontinuity (reverse shock). 
The thermal parameters of the cool region at $t = 0.5$, 1.0, and 1.4~yr
are plotted as crosses in Fig.~\ref{fig:teq}.

Beyond $t=1.4$~yr, as the cloud cools further, its thermal energy as a 
fraction of its total energy (thermal plus kinetic) approaches the 
round-off error of our calculation ($\sim 10^{-4}$), and we
cannot follow its evolution past this time. There is also the well-known
problem of the `eating away' of the edges of hot material. This affects
all {\mbox numerical} schemes, and results from the numerics smearing the 
temperature gradient and placing cells at intermediate temperatures, which
then undergo high cooling (often the line-cooling bump at $T \sim 10^{5}$~K). 
In this fashion cool regions can rapidly grow as hot regions are 
`eaten away'.

However, it is clear that the gas will continue to cool, and eventually 
reach $T \approx 2 \times 10^{4}$~K, as the following argument demonstrates. 
With $t_{c} << t_{d}$ the gas cools isochorically so 
$p \propto T \propto 1/\Xi$. At $t=1.4$~yr, the cloud
is at a temperature $T = 7.0 \times 10^{5}$~K and an ionization 
{\mbox parameter} $\Xi = 10.8$. It therefore moves approximately {\mbox along} 
a line of
slope -1 in Fig.~\ref{fig:teq}, and encounters the thermal {\mbox equilibrium}
curve at $\Xi \approx 40$ and $T \approx 2 \times 10^{5}$~K. This point is 
thermally stable and the rapid decrease in the temperature is brought 
to a halt. The temperature of the cool region, $T_{c}$, then remains 
relatively constant, and ajusts only for changes in the value of the 
ionization parameter, $\Xi_{c}$, which occur on a sound-crossing 
timescale as the surrounding post-shock material gradually repressurizes 
the cool gas to its level. The sound
crossing time within the shocked ejecta is short compared to the dynamical
time of the remnant, so $\Xi_{c} \rightarrow \Xi_{ss}$, where $\Xi_{ss}$ is
the ionization parameter of the surrounding shock material ($\Xi_{ss}
\approx 2-3$ in the shocked ambient gas). There is therefore no immediate 
danger of $\Xi_{c}$ 
increasing to the point that the cooled gas begins to reheat towards the 
Compton temperature. Thus $\Xi_{c}$ approaches $\Xi_{ss}$ and the  
cooled region moves along the thermal equilibrium curve to the left. 
The density of the region, $n_{c}$, increases during this process until 

\begin{equation}
\label{eq:ncloud}
n_{c} = \frac{T_{c'}}{T_{c}} n_{c'},
\end{equation}

\noindent where $n_{c'}$ ($T_{c'}$) is the density (temperature) of the 
cooled material at the end of our simulation. From the results in
Fig.~\ref{fig:clump_form}, we obtain $n_{c} \approx$ few 
$\times 10^{10} \; {\rm cm^{-3}}$, in good {\mbox agreement} 
with inferred values of the density of the HIL region from observations.
We can also estimate the resultant {\mbox hydrogen} column of the region: at
$t = 1.4$~yr, its thickness is approximately $2 \times 10^{13}$~cm (the
numerical resolution of our calculation is about half this), and its 
density is $\approx 3 \times 10^{8} \; {\rm cm^{-3}}$, which for solar 
abundances results in a column density of $\approx 10^{22} \; {\rm cm^{-2}}$, 
again in agreement with observations of the HILs. The radius of the remnant 
is $\approx 7.5 \times 10^{-3}$~pc at this stage, 
whilst the expansion speed of the contact discontinuity (and thus
the velocity of the cool region) is 
$\approx 4 \times 10^{8} \; {\rm cm \; s^{-1}}$.

Observations of the HILs have revealed that the optical depth is less than 
but of order unity, and that $N_{H} \approx 10^{22}\;{\rm cm^{-2}}$. 
The observed range of the ionization {\mbox parameter} of the HIL region is 
$0.3 \ltsimm \Xi \ltsimm 2$ (Kwan \& Krolick \cite{KK1981}), which is 
somewhat higher than that inferred for the LIL region. The systematic 
blue-shift of the HILs with respect to the LILs is interpreted as the 
HILs forming in a region with bulk outflow (possibly inflow) and that 
the red-shifted emission is obscured. Line widths, which reflect local 
gas velocities, are typically several $10^{3} \kmps$. Hence our 
model results are in harmony with the temperature, 
ionization parameter, column density, and velocity dispersion of the
observed HIL BELR clouds.

We now check to see if our results are in agreement with the two
conditions which must be satisfied for the shocked gas to cool down to
$T \approx 2 \times 10^{4}$~K, namely $t_{cool} \ltsimm t_{dyn}$ and
$\Xi)_{t_{dyn}} \ltsimm \Xi_{crit}$. For $n=12$, $\delta=0$, 
E~=~$10^{51}$~ergs,
$M_{ej} = 10 \;\Msol$, $R_{2}/R_{c} = 0.974$, A~=~0.19 and 
$\rho_{w} = 10^{-18}$~${\rm g\;cm^{-3}}$, we obtain
from Eq.~\ref{eq:t_dyn} $t_{dyn}$~=~7~yr. Assuming solar abundance for 
the ejecta, $\mu_{s} = 0.61$,
and $t_{cool} = 3$~yr. Hence $t_{cool} < t_{dyn}$ as required. The fact
that the cooled region first forms at $t \approx 1.4$~yr indicates that
we are underestimating the cooling rate at lower temperatures where 
line-cooling is dominant. For these parameters, we further obtain
from Eq.~\ref{eq:v_core} $v_{c} = 3610$~$\kmps$, and from 
Eq.~\ref{eq:xi_shk_t_dyn} $\Xi_{shk})_{t_{dyn}} = 6.3$, and thus Model~A
also satisfies the requirement that $\Xi_{shk})_{t_{dyn}} \ltsimm 20$.
In reality the ejecta will be enriched in metals, which will reduce
the above estimate of $t_{cool}$.

\subsection{Exploration of associated parameter space}
\label{sec:explore_param_space}

\subsubsection{Variation of $n_{w}$, $r_{pc}$, and $L_{ion}$}
\label{sec:var1}
Since $n_{w}$, $r_{pc}$, and $L_{ion}$ all influence $\Xi_{shk}$ it is
impossible to discuss the effect of a variation in one without also
considering the impact of the others. This inevitably leads to some
complexities but we attempt to keep our discussion as simple as possible.
 
We can first consider the effect of varying the assumed ambient density.
Since the equilibrium temperature is a function of $\Xi$, we wish
to keep this value constant for our initial investigation. As
$\Xi_{w} \propto L_{ion}/(n_{w} r_{pc}^{2})$, this implies a variation
in flux from the central engine, which can {\mbox obviously} be interpreted as
a variation in the ionizing luminosity $L_{ion}$ with $r_{pc}$ fixed, 
or as a variation in the distance $r_{pc}$ between the remnant and the central 
engine with $L_{ion}$ fixed, or a suitable variation in both $L_{ion}$
and $r_{pc}$.

\begin{figure*}
\begin{center}
\psfig{figure=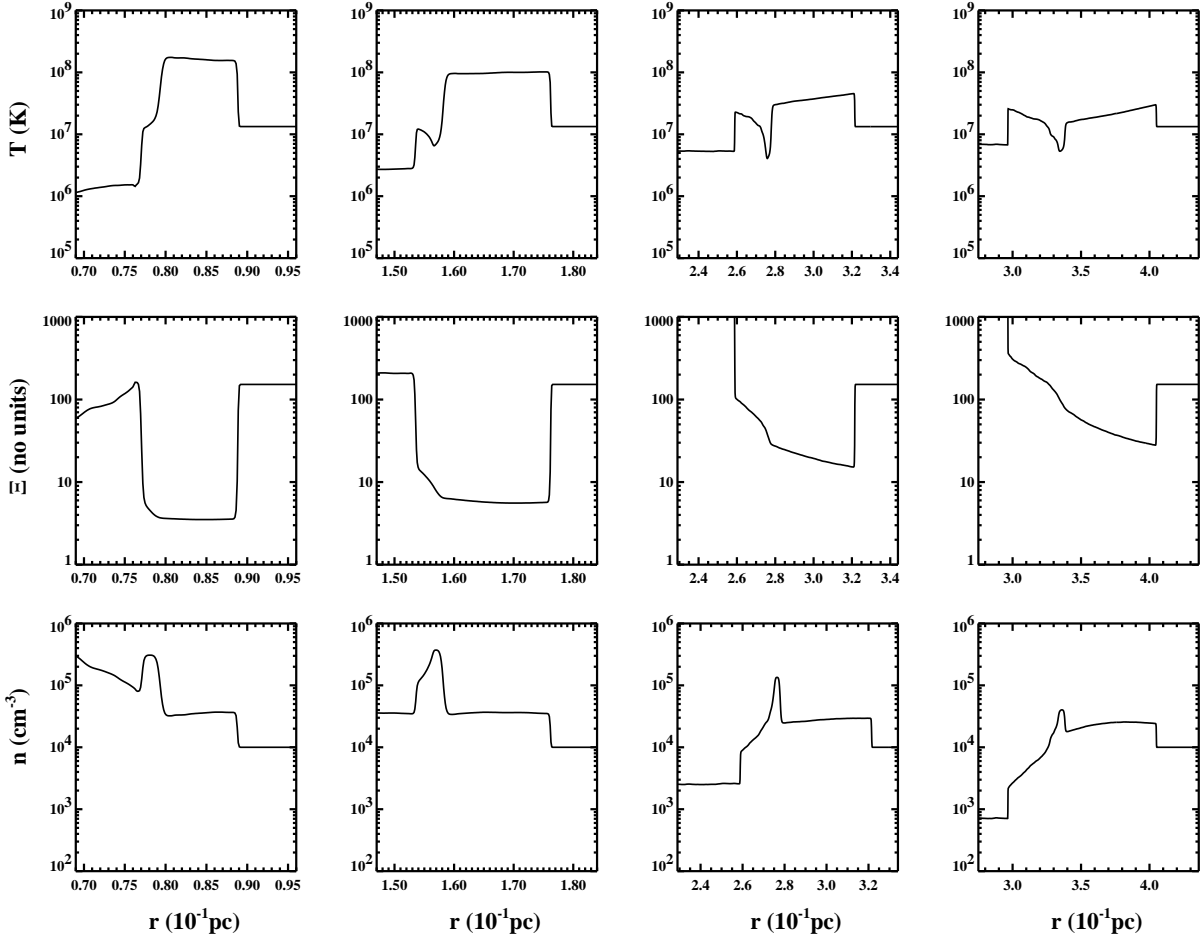,width=16.0cm}
\end{center}
\caption[]{As Fig.~\ref{fig:clump_form} (Model~A) but for a constant 
density medium with $n_{w} = 10^{4} \;{\rm cm^{-3}}$, an ionizing AGN 
luminosity of $10^{46} \ergps$, and a distance of 1~pc to the central engine
(Model~B). $\Xi_{w}$ is again $\approx 150$. From left to right the 
corresponding times are 20~yr, 50~yr, 120~yr and 180~yr.
In contrast to Model~A, cool clouds have no opportunity 
to form since the interaction of the ejecta core with the reverse shock leads
to a rapid depressurization of the shocked gas.}
\label{fig:clump_noform}
\end{figure*}

In Fig.~\ref{fig:clump_noform} we show the results for a simulation 
with $\Xi_{w} \sim 150$ (matching the earlier simulation of Model~A shown in 
Fig.~\ref{fig:clump_form}) but with a lower ambient density 
($n_{w}$~=~$10^{4} \; {\rm cm^{-3}}$) and AGN flux (\eg 
$L_{ion} = 10^{46} \ergps$ with the remnant 1~pc distant). 
We call this Model~B. As a result of the lower ambient density, it has a 
longer evolution time ($t_{dyn} = 32$~yrs) and the remnant expands further 
than in Model~A. Relative to Model~A, the lower flux from the central engine 
also increases the cooling timescale of the shocked gas 
($t_{cool} = 300$~yrs). Although the ionziation parameter of the shocked gas 
at $t = t_{dyn}$ is less than $\Xi_{crit}$ ($\Xi)_{t_{dyn}} = 6.4$, as for
Model~A), because $t_{cool} > t_{dyn}$ we do not expect any of
the shocked material to cool to $T \approx 2 \times 10^{4}$~K,

Once the reverse shock reaches the core radius
at $t = 32$~yr, the solution swiftly diverges from a 
self-similar form (despite the cooling of the post-shock material, many 
aspects such as the radius and velocity of the shocks are not too far 
removed from such an evolutionary form prior to this point). At this time the
coolest shocked gas is at $T \approx 7 \times 10^{6}$~K. The subsequent
acceleration of the reverse shock towards the 
centre of the remnant rapidly de-pressurizes the shocked region, increasing
the ionization parameter of the shocked gas, and ultimately leading to its
reheating towards the Compton temperature. Thus, by simply adjusting the
density of the ambient medium, whilst keeping all other parameters constant,
we have shown that a cool region may not form if the effective ram pressure
of the ambient medium is not sufficiently high.

We consider whether there is also a mechanism which prevents the formation 
of cool regions in SNRs once the density of the ambient material increases 
passes some {\mbox limit}. This question can be addressed in the 
following way. 
As $n_{w}$ increases, $L_{ion}/r_{pc}^{2}$ must similarly increase to 
maintain a given value for $\Xi_{w}$. For a given $L_{ion}$ this means a 
reduction in $r_{pc}^{2}$ (which might be consistent with an increase 
in $n_{w}$). A higher value of $n_{w}$ leads to a faster evolution of the 
remnant, which given that the cooling timescale of the shocked gas also 
shortens, does not prevent the possibility of
cool regions forming. However, physically it does mean that the lifetime 
of the cool regions is also very short, {\mbox since} the evolved time 
between their 
formation and their destruction, which presumably occurs shortly after
the interaction of the ejecta core with the reverse shock, is
short too. Hence at some value of $n_{w}$, the contribution to the 
BELR emission will be on such short timescales that it will become an
unimportant component of the overall emission. 

We note that it is unlikely that the ejecta density actually has such a 
sharp cut in its gradient. However, once the ejecta passing through the 
reverse shock starts deviating from an $r^{-12}$ profile the volume 
of the shocked region will
start expanding with an inevitable rise in the ionization parameter. This
would occur at an earlier time relative to the sharp cut-off case and 
would tend to reduce the contribution of mass to the BELR by this model.
Assuming a constant value of $\Xi_{w}$ and 
$L_{ion}$, this argument also introduces a lower bound to
the value of $r_{pc}$ at which a cool region can form.

Conversely, for a given $r_{pc}$, an increase in $n_{w}$ and constant
$\Xi_{w}$ means a corresponding increase in $L_{ion}$. In a similar
way to the above argument, this introduces an {\mbox upper} bound to the 
value of
$L_{ion}$ at which a cool region can {\mbox form}. These conditions together 
place upper bounds on the values of $n_{c}$ and $v_{c}$ inferred from 
observations. 

\subsubsection{Variation of $\Xi_{w}$}
We now consider the effect of a different wind ionization parameter
on the possible formation of a cool BELR. In our
simulations so far we have deliberately specified a value of $\Xi_{w}$
which is high enough for the ambient medium to exist at the Compton 
temperature, but which is at the same time almost as low as we could 
make it, providing us with the best opportunity to form a cool region
(through a low value of $\Xi_{shk}$ \cf Eq.~\ref{eq:ionp_shk}).
If $\Xi_{w}$ was increased in value (and given that $T_{w}$ is almost
constant for large values of $\Xi_{w}$) we would require a higher
relative velocity, $\omega$, to obtain a given value of $\Xi_{shk}$
(\cf Eq.~\ref{eq:ionp_shk}). For a given remnant age and $n_{w}$, and
assuming a static medium, a higher value of $\Xi_{w}$ gives a higher
value of $\Xi_{shk}$, and since $\Xi_{shk}$ gradually increases with
time as the expansion velocity of the remnant slows, this effectively
reduces the `window' of opportunity where $\Xi_{shk}$ is low enough
for $T_{eq} \sim 10^{4}$~K. This in turn places a limit on
the maximum cooling timescale for the shocked gas, which for a given value
of $\Xi_{w}$, feeds back into a lower limit on the ambient density, 
$n_{w,min}$. If $n_{w} < n_{w,min}$, the shocked gas 
will not have enough time to cool to $\sim 10^{4}$~K before $\Xi_{shk}$
becomes too large.

\subsubsection{Variation of $\omega$}
\begin{figure*}
\begin{center}
\psfig{figure=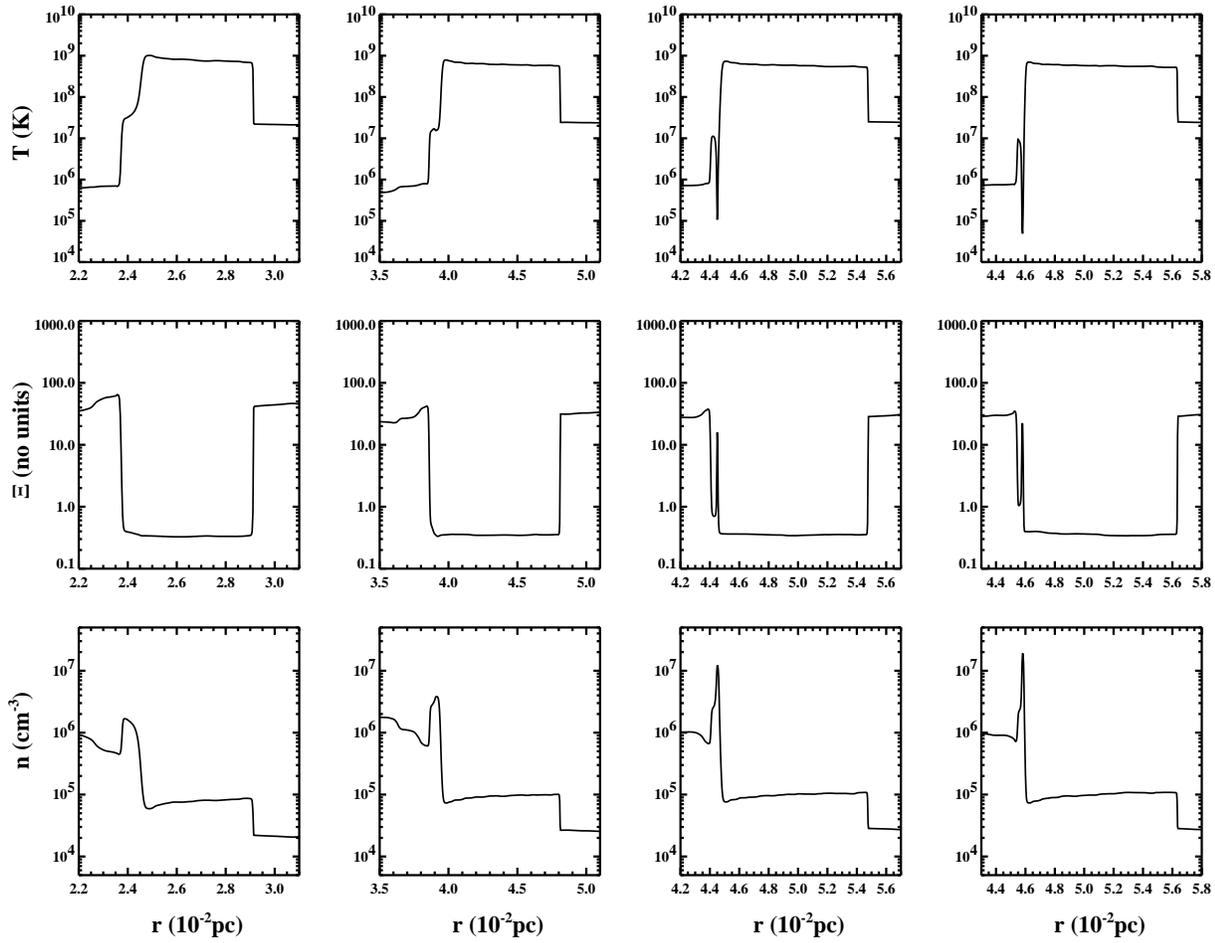,width=16.0cm}
\end{center}
\caption[]{As Fig.~\ref{fig:clump_noform} (Model~B) but in this
simulation (Model~C) the ambient medium has a flow velocity towards the 
remnant of $3,000 \kmps$. From left to right the corresponding times are
5~yr, 10~yr, 12~yr and 12.5~yr. In contrast to the results for a
stationary medium, cool clouds {\em can} form in this case. Note also the
more rapid evolution of the remnant with respect to the case shown in 
Fig.~\ref{fig:clump_noform}.}
\label{fig:clump_form_agn46_u3e8}
\end{figure*}

If the medium surrounding the remant has some bulk {\mbox velocity} of its own,
it is possible to obtain the same value of $\Xi_{shk}$ at the same remnant
age, for a higher value of $\Xi_{w}$. On the side of 
the remnant facing the oncoming `AGN wind' the expansion of the remnant is
inhibited, leading to an increased value for $v_{rel}$ 
(\cf Eqs.~\ref{eq:xi_shk} and~\ref{eq:t_shk}) or alternatively $\omega$
(\cf Eq.~\ref{eq:ionp_shk}). In turn this reduces the critical density of the
ambient medium needed for a cool region to form, $n_{w,min}$. In 
Fig.~\ref{fig:clump_form_agn46_u3e8} we show the results for Model~C, a 
remnant of a type-II supernova expanding into a {\mbox constant} 
density medium with 
$n_{w} = 10^{4} \;{\rm cm^{-3}}$, $\Xi_{w} \approx 150$, and a flow 
velocity of $3,000 \kmps$ (\cf Williams \etal \cite{WBP1999}) towards the
remnant. These results are indicative of the remnant structure on the 
upstream side (since the shocked region is thin with respect to the radius 
of the {\mbox remnant}). The entire area is again immersed in a 
radiation field 
appropriate to a QSO of luminosity $10^{46} \ergps$ with the central
engine at a distance of 1~pc. Fig.~\ref{fig:clump_form_agn46_u3e8} shows
that this time a cool region {\em can} form. In this case the ionization 
parameter of the cool gas is expected to be $\approx 0.5$ and its density
to be $n \approx 10^{9}\;{\rm cm^{-3}}$. Its column density is
$\approx 6 \times 10^{21}\;{\rm cm^{-2}}$ and its speed is 
$\approx 2580 \kmps$. These values are again in good agreement with 
observations.

\subsubsection{Constraints on the parameter range}
\label{sec:constraints}
The previous sections show that there are two broad constraints on the 
parameter range required for
cool regions to form in SNRs. First, the post-shock ionization parameter
needs to be low enough for the equilibrium temperature of this region to
approach $\approx 10^{4}$~K. This requires that the SNR evolves in
a dense medium (Model~A (Fig.~\ref{fig:clump_form}) vs. Model~B 
(Fig.~\ref{fig:clump_noform})) but at the same time is not too close to 
the central engine (\cf Eq.~\ref{eq:xi_shk_t_dyn}).
Whilst tending to decrease the evolutionary timescale of the remnant, a
higher effective confining pressure increases the density of the 
shocked gas leading to shorter cooling timescales. Thus although a high
ambient density does not inhibit the formation of a cool region, at some
level it results in such rapid evolution that these regions are destroyed
without making any substantial contribution to the HIL emission.

The distance to the central engine influences the value of 
$\Xi_{shk}$ through the ionzing flux (Eq.~\ref{eq:xi_shk})
Therefore there is an upper limit to $\Xi_{w}$ after which not even 
SNe in the densest environments will be able to produce cool regions.
If the ambient medium is in motion relative to the
remnant, the parameter space over which cool regions can form is extended
(Model~B (Fig.~\ref{fig:clump_noform}) vs. Model~C 
(Fig.~\ref{fig:clump_form_agn46_u3e8})).

The second constraint is that the ambient medium needs to be dense enough 
for the cooling timescale of the shocked gas to be less than the age of 
the remnant at the time when the reverse shock reaches ejecta which do 
not have a strongly radially dependent density. If this is not the case no
cool regions can form, as illustrated by Model~B 
(Fig.~\ref{fig:clump_noform}).

\section{Discussion}
\label{sec:conclusions}
The results in Sec.~\ref{sec:results} demonstrate that it is possible to
cool shocked supernova ejecta down to $T \sim 10^{4}$~K in the inner 
regions of a QSO. Although our results differ from the original proposals 
of Dyson \& Perry (\cite{DP1982}) and Perry \& Dyson (\cite{PD1985}), which
were for the shocked ambient medium to cool, the resulting cool
gas nevertheless has properties (densities, column depths, velocities and 
ionization parameters) compatible with those inferred for gas 
emitting the high ionization lines in QSOs. We have additionally shown 
from three separate calculations that this model is able to cover the 
parameter {\em range} associated with observations of BELRs, and have 
highlighted scenarios which are outside of this range and preclude 
the formation of cool regions.

We now compare our results to some of the observational correlations
that have been determined. Korista \etal (\cite{K1995}) noted that the blue and
red wings respond faster to continuum changes than the core. If one
assumes that the expansion velocity of the supernova remnant is the 
dominant velocity component, this observation can be explained with
our model in the following way. Remnants which are surrounded by the
highest ambient densities {\mbox evolve} most rapidly, 
and therefore form cool regions
with the highest expansion velocities. If the ambient density increases
towards the central engine, there will be a {\mbox tendency} for remnants 
in these
regions to form cool regions with, on average, higher velocities. Being
close to the central engine, these regions respond more 
rapidly to any continuum changes, as the observed correlation requires.  
Even if $v_{exp}$ is not the dominant component, the above
{\mbox argument} implies that our assumed model {\mbox nevertheless} 
provides a statistical bias in this direction.

There now appears to be conclusive evidence that the higher ionization lines 
respond faster than lower ionization lines to changes in the continuum 
variability (see references in Fromerth \& Melia \cite{FM2001}). 
If the LIL are formed in an accretion disc close to the central 
engine, this implies that the thermal timescale in the disc is longer than
the corresponding timescale for the HIL. Alternatively the delay could be
explained if the bulk of the LIL heating is due to back-scattered X-rays 
from the gas responsible for the HILs and the general ambient medium 
(see Collin-Souffrin \etal \cite{CDMP1988}). Therefore, our model is
consistent with these observations.

Our calculations show that it is the shocked {\em ejecta} which cool to form
the HILs, and that once the ejecta core reaches the reverse shock, the
shocked region rapidly depressurizes. Since this presumably leads to
the destruction of any cool regions, it seems sensible to suppose that
for the remnant parameters we have chosen, an upper {\mbox limit} to the HIL
mass would be $2.5 \;\Msol$ per remnant. With further appropriate assumptions
we can then estimate the required supernova rate needed to sustain the
mass in the gas responsible for the HIL. With the assumption that
each remnant can eventually cool $\sim 2.5 \;\Msol$ of gas, and that the
lifetime of the cool gas in the remnant is of order 10~yr, to 
sustain $\sim 25 \; \Msol$ of gas in the HIL would require a supernova rate of 
$\sim 2 \;{\rm yr^{-1}}$. Although our estimates are very uncertain,
this rate is acceptable:
certainly for high luminosity QSOs, supernova rates 
$\sim 10 \; {\rm yr^{-1}}$ are conceivable (Terlevich \etal \cite{TTFM1992}).
A less steep density distribution of 
the ejecta envelope (if $n < 12$ more mass would be available per SN), 
and the suppression of dust formation in intermediate-mass
AGB stars in the BELR region (which may reduce the minimum
zero-age main sequence mass required for supernovae; Hartquist \etal 
\cite{H1998}) are {\mbox two}
additional possibilities which could further reduce our estimate of the 
required supernova rate. Additionally, it is possible that conditions
exist for the swept-up ambient material to also cool to temperatures
appropriate for the HIL, as the cooling time estimated by Eq.~\ref{eq:t_cool}
is {\mbox temperature} independent. The fact that the reverse shocked material 
cools first in our simulations is simply due to enhanced cooling by 
line emission starting earlier due to its lower initial temperature. 
In Model~A where $t_{cool} < t_{dyn}$ we therefore might expect the 
swept-up ambient material to form HIL gas as well. At $t=t_{dyn}$, 
the forward shock is estimated to be at a radius of 0.03~pc and to have 
swept up $1.6 \Msol$ of ambient gas. This gas could therefore be a 
significant contributer to the total HIL emission.

In this first paper we have been solely interested in the question of 
whether cool regions could form from supernova shocks given that they 
are bathed in the hard radiation flux of the central engine.
For simplicity we restricted our modelling to the simplest 1D approach, and
assumed solar metallicites in our calculations. The fact that the 
{\mbox ejecta} (and also the swept up medium) may be {\mbox responsible} 
for the HIL 
emission warrants a careful consideration in future models. We will 
also perform calculations on 2D axisymmetric hydrodynamic grids. 

Another future goal is the inclusion of the effect of the QSO radiation field
on the {\em dynamics} of the remnant. At high temperatures, the effective
cross-section of gas is the Thomson scattering cross-section, $\sigma_{T}$.
Once the gas begins to cool, the effective cross-section, $\bar{\sigma}$, 
increases, enhancing the radiative driving. As noted by Williams 
\etal (\cite{WBP1999}), $\bar{\sigma}/\sigma_{T}$ increases to roughly 
2 at $10^{6}$~K, 40 at $10^{5}$~K,
and to $\geq 10^{4}$ when the gas has fully cooled (see also 
Arav \& Li (\cite{AL1994}) and Arav \etal (\cite{ALB1994})). 
For gas cooler than approximately $5 \times 10^{4}$~K,
it is reasonable to take the radiative acceleration as 
$g = 7 \times 10^{14} \; \rho$ (R\"{o}ser \cite{R1979}; Dyson \etal 
\cite{DFP1981}). We expect that
substantial radiative driving will critically alter the
dynamics of the interaction between the supernova ejecta and the ambient
medium (\eg Falle \etal \cite{F1981}; Williams \cite{W2000}).

Although we have chosen to model the interaction of supernova ejecta with the
ambient medium, it is possible that a wind from a group 
of early-type stars may also provide the necessary conditions for the
formation of cool regions. This interaction may be more relevant 
in the nuclei of Seyfert galaxies, since supernova explosions
will evacuate all but the most tightly bound gas in them (Perry \& Dyson
\cite{PD1985}). Finally it is clear from our models that whilst the
supernova-QSO wind interaction is conceptually simple, the BELR
is likely to be a very complicated region in practice.

\begin{acknowledgements}
JMP would like to thank PPARC for the funding of a PDRA position. 
We would like to thank R.J.R. Williams and R. Coker for many helpful 
conversations and T. Woods for the use of his cooling and heating tables.
This research has made use of Nasa's Astrophysics Data System Abstract 
Service. We would also like
to thank an anonymous referee whose suggestions improved this paper.
\end{acknowledgements}

\appendix
\section{Derivation of $t_{dyn}$ and $\Xi_{shk}$}
\label{app:appendix}
From Chevalier (\cite{C1982}), the radius of the contact {\mbox discontinuity} 
for a supernova remnant expanding into a stationary, constant density ambient
medium ($\rho_{w}$~=~$n_{w} \mu_{w} \massh$, where $\mu_{w}$ is the average 
mass per {\mbox particle}), is 

\begin{equation}
\label{eq:r_cd}
R_{c} = \left(\frac{A g^{n}}{n_{w} \mu_{w} \massh}\right)^{1/n} 
t^{(n-3)/n},
\end{equation} 

\noindent where $A$ is a constant determined by the
steepness of the ejecta profile ($n$). The radius of the reverse shock is 

\begin{equation}
\label{eq:r_rs}
R_{2} = \frac{R_{2}}{R_{c}} R_{c},
\end{equation}

\noindent where $R_{2}/R_{c}$ is also a function of $n$. For $n=12$, 
Chevalier's Table~1 gives $A = 0.19$ and $R_{2}/R_{c} = 0.974$. The velocity
of the pre-shock ejecta at the reverse shock is 

\begin{equation}
\label{eq:v_ej_rs}
v_{ej})_{R_{2}} = R_{2}/t.
\end{equation}

\noindent The dynamical time (\ie the time at which the ejecta core reaches
the position of the reverse shock) can be obtained from the ratio

\begin{equation}
\label{eq:t_dyn_ratio}
\frac{t_{dyn}}{t} = \left(\frac{v_{c}}{v_{ej})_{R_{2}}}\right)^{-n/3},
\end{equation}

\noindent where $v_{ej})_{R_{2}}$ is the velocity of the pre-shock ejecta
at the reverse shock at time $t$. Substituting 
Eqs.~\ref{eq:v_core},~\ref{eq:g_n} and~\ref{eq:v_ej_rs} we obtain

\begin{eqnarray}
\label{eq:t_dyn}
t_{dyn} & = & \left(\frac{A}{4 \pi (n-\delta) n_{w} \mu_{w} 
\massh}\right)^{1/3} \left(\frac{R_{2}}{R_{c}}\right)^{n/3} \nonumber \\ 
& & \hspace{20mm} \times 
\frac{[2(5-\delta)(n-5)E]^{-1/2}}{[(3-\delta)(n-3)M_{ej}]^{-5/6}}.
\end{eqnarray}

Since the shocked gas is initially at very high temperatures, we make the
assumption that the dominant cooling during this interval is inverse 
Compton. The radiative losses (${\rm erg\;cm^{-3}\; s^{-1}}$) 
are given by

\begin{equation}
\label{eq:edot_ic}
\dot{E} \approx 2.2 \times 10^{7} \frac{L_{47}}{r_{pc}^{2}} T_{8} \; \rho
\end{equation}

\noindent (Perry \& Dyson \cite{PD1985}). The cooling timescale is 
independent of density and temperature, and is given by

\begin{equation}
\label{eq:t_cool}
t_{cool} = \frac{E}{\dot{E}} \approx 18 \frac{r_{pc}^{2}}{\mu_{s} 
L_{47}} \;\; {\rm yrs},
\end{equation}

\noindent where $\mu_{s}$ is the average mass per particle for the shocked
material. For the post-shock gas to cool down to approximately 
$T = 2 \times 10^{4}$~K, we require $t_{cool} \ltsimm t_{dyn}$. 

We now derive an expression for the evolution of the ionization parameter 
of post-shock gas cooled back to {\mbox equilibrium} with the radiation field, 
$\Xi_{s}$. From Perry \& Dyson (\cite{PD1985}) we have

\begin{equation}
\label{eq:xi_shk}
\Xi_{shk} = \frac{L_{ion}}{4 \pi r^{2} n_{s} k T_{s} c} 
= 2.0 \times 10^{14} \frac{L_{ion,47}}{r_{pc}^{2} n_{s} T_{s}},
\end{equation}

\noindent where $n_{s}$ is the number density of the shocked material,
$T_{s}$ is the immediate post-shock temperature of the material and is
given by

\begin{equation}
\label{eq:t_shk}
T_{s} = \frac{3}{16} \frac{\mu_{s} \massh}{k} v_{r}^{2},
\end{equation}

\noindent and $v_{r}$ is the velocity of the pre-shock ejecta in the 
frame of the reverse shock. The velocity of the reverse shock is

\begin{equation}
\label{eq:v_rs}
v_{R_{2}} = \frac{R_{2}}{R_{c}} v_{cd} = \frac{n-3}{n} \frac{R_{2}}{R_{c}}
\left(\frac{A g^{n}}{n_{w} \mu_{w} \massh}\right)^{1/n} t^{-3/n},
\end{equation}

\noindent so using Eq.~\ref{eq:v_ej_rs} we obtain

\begin{eqnarray}
\label{eq:v_rel}
v_{r} & = & \left[1 - \left(\frac{n-3}{n}\right)\right] \frac{R_{2}}{R_{c}}
\left(\frac{A}{4 \pi n_{w} \mu_{w} \massh (n-\delta)}\right)^{1/n} 
\nonumber \\
 & & \hspace{3mm} \times 
\frac{[2(5-\delta)(n-5)E]^{(n-3)/2n}}{[(3-\delta)(n-3)M_{ej}]
^{(n-5)/2n}} t^{-3/n},
\end{eqnarray}

\noindent which at $t=t_{dyn}$ simplifies to

\begin{equation}
\label{eq:v_rel_t_dyn}
v_{r})_{t_{dyn}} = \left[1 - \left(\frac{n-3}{n}\right)\right] v_{c}.
\end{equation}

\noindent Combining Eqs.~\ref{eq:xi_shk},~\ref{eq:t_shk} 
and~\ref{eq:v_rel_t_dyn}, and noting that as it is the reverse shock 
which cools, $n_{s} = 4 n_{w} \rho_{2}/\rho_{1} = 
4 \rho_{w} \rho_{2}/(\rho_{1} \mu_{w} \massh)$, we obtain

\begin{equation}
\label{eq:xi_shk_t_dyn}
\Xi_{shk})_{t_{dyn}} \approx \frac{1}{27} \frac{\rho_{1}}{\rho_{2}}
\frac{L_{ion,47}}{r_{pc}^{2} n_{w} \mu_{w} \massh}
\left(\frac{1}{\left[1 - \left(\frac{n-3}{n}\right)\right] v_{c}}\right)^{2}.
\end{equation}

\noindent Values for $\rho_{1}/\rho_{2}$ are given in
Chevalier (\cite{C1982}).

\end{document}